\documentstyle[preprint,eqsecnum,aps]{revtex}

\begin{document}
\input epsf

\count255=\time\divide\count255 by 60 \xdef\hourmin{\number\count255}
  \multiply\count255 by-60\advance\count255 by\time
  \xdef\hourmin{\hourmin:\ifnum\count255<10 0\fi\the\count255}

\draft
\preprint{WM-00-105}

\title{Predictions for Decays of Radially Excited Baryons}

\author{Carl E. Carlson\footnote{carlson@physics.wm.edu}
 and  Christopher D. Carone\footnote{carone@physics.wm.edu}}

\vskip 0.1in

\address{Nuclear and Particle Theory Group, Department of
Physics, College of William and Mary, Williamsburg, VA 23187-8795}

\vskip .1in
\date{May, 2000}
\vskip .1in

\maketitle
\tightenlines

\begin{abstract}
We consider the strong decays of the lowest-lying radially excited 
baryons, in SU(6) language the states comprising the first 
excited {\bf 56}-plet. Assuming a single-quark decay approximation, 
and negligible configuration mixing, we make model-independent predictions 
for the partial decay widths to final states with a single meson.  Masses
of unobserved states are predicted using results from large-$N_c$
QCD, and the momentum dependence of the one-body decay amplitude is
determined phenomenologically by fitting to observed decays, so that 
the baryon spatial wave functions are not assumed.  We point out
that comparison of these predictions to experiment may shed light on
whether the Roper resonance can be consistently interpreted as a 
three-quark state.
\end{abstract}

\thispagestyle{empty}

\newpage
\setcounter{page}{1}

\section{Introduction} \label{sec:intro}

In this note, we consider the strong decays of the lowest-lying
radially excited baryons.  These are the states comprising
the first excited {\bf 56}-plet, in SU(6) language. We make 
predictions for the decay widths of these states that are independent 
of any specific model for the binding potential or spatial wave functions.  

A reason for taking up studies of excited baryons now
is that we anticipate new results from the CLAS Collaboration
at the Thomas Jefferson National Accelerator Facility. 
Not all the states in the radially excited {\bf 56}-plet are yet
discovered, and many of the measured decay widths of the
observed states have large uncertainties.  A set of
predictions for the decay widths of unobserved states will be of
great use to experimenters. Of necessity, we also provide
predictions for the masses of the unobserved states.
To this end, we present a derivation of the G\"ursey-Radicati mass 
formula~\cite{GR} from large-$N_c$ QCD that we believe has not appeared
explicitly in the literature. 

Another and more specific reason for our analysis is to shed
light on the Roper resonance, the $N(1440)$.  The Roper is the first
excitation of the nucleon with the same $J^P = (1/2)^+$
quantum numbers.  The most direct explanation of the Roper is as 
a three-quark radially excited state. However, certain of its apparent  
peculiarities have led to plausible suggestions that it may be a 
hybrid state~\cite{hybridroper} (a state whose lowest significant 
Fock component is three quarks plus gluonic excitations) or that it 
may be a cross section enhancement that does not correspond to any
resonance~\cite{dynamicalroper}.  An argument against interpreting the 
Roper as a three-quark state has been that its calculated mass is too 
high in quark models with one-gluon-exchange, {\em i.e.} spin-color,  
interactions~\cite{isgurkarl1}.  Recently, it has been found that 
the observed mass is perfectly consistent with the three-quark picture 
if mass splittings are due to spin-flavor interactions 
instead~\cite{glozmanriska}.  Moreover, there is evidence of a Roper 
signal from a lattice calculation using a three-quark source operator, 
although uncertainties in the mass determination as yet preclude saying 
if it is lighter than its negative parity counterpart~\cite{sasaki}. In 
the present work, we assume the Roper resonance is a three-quark state 
embedded in the excited {\bf 56}-plet; confirmation of the decay widths 
predicted here will therefore support the three-quark interpretation of 
this state.

The value of a model-independent analysis should be clear:
We are free of any assumptions regarding the binding potential 
or quark spatial wave functions. We do make assumptions about 
the dominance of single-quark operators and about configuration 
mixing that we now consider in turn.

First, we assume that the decays of interest proceed via a single quark 
interaction vertex.  We may write the decay operator in 
the same SU(6) tensor product space in which we define the baryon 
spin-flavor wave functions:
\begin{equation}
{\cal H}_{{\rm eff}} \propto G_*^{ia} k^i \pi^a  .
\label{eq:op}
\end{equation}
Here $G_*^{ia}= S_*^i T_*^a$ is an SU(6) generator, where $S_*^i$ and $T_*^a$
are the generators of SU(2) and SU(3), respectively, $k^i$ is the
meson three-momentum, and  $\pi^a$ represents a meson field operator.
The asterisk in Eq.~(\ref{eq:op}) indicates that that the operator acts on the
spin-flavor wave function of the excited quark.  Here we adopt a
convenient picture of the baryon state that follows from large-$N_c$
QCD, namely, that the baryon consists of a `core' of $N_c-1$ ground 
state quarks generating a collective potential in which a single 
quark is excited.  Note that the complete symmetry of the {\bf 56}-plet 
spatial wave functions implies that matrix elements of the operator 
in Eq.~(\ref{eq:op}) are proportional to matrix elements of the same 
spin-flavor operator summed over all the quarks in the baryon state,
\begin{equation}
G^{ia} \equiv \sum_{{\rm quarks}\,\,\, \alpha} S^i_\alpha T^a_\alpha \, .
\end{equation}
Thus, the operator in Eq.~(\ref{eq:op}) works equally well in 
parameterizing the decays of radially excited baryons for $N_c=3$, where 
the  state may be described more realistically as a collective excitation 
of all three quarks.  Matrix elements of this operator may be written
\begin{equation}
\langle \Psi(B_f,\pi^a) | {\cal H}_{{\rm eff}} | \Psi(B_i) \rangle
    =  f(k) k^j \langle B_f|  G^{ja} | B_i  \rangle
\label{eq:mop}
\end{equation}
where $k = |\vec k|$ and  $f(k)$ is a function that parameterizes 
the momentum dependence of the amplitude.  This includes momentum
dependence that originates from the underlying quark-level vertex,
as well as from the overlap of the baryon spatial wave functions.
The baryon states shown represent only the spin-flavor part of the
wave function, and are nonrelativistically normalized.  Use of the 
single-body decay approximation is strongly motivated by its success in 
describing decays of the orbitally excited, {\bf 70}-plet baryons, both 
in the algebraic SU(6) approach~\cite{algebraic}, as well as more recent 
large-$N_c$ effective field theory 
analyses~\cite{largen1,largen2,largen3,largen4}.  In the latter case, 
both one- and two-body operators have been 
shown to contribute to {\bf 70}-plet decay matrix elements at leading 
order in the $1/N_c$ expansion.  Nonetheless, comparison to data for both 
strong decays~\cite{largen2} and photoproduction amplitudes~\cite{largen3}
reveals that the two-body operators are phenomenologically irrelevant.  It 
is not unreasonable to assume that similar dynamics may play a role in the 
strong decays of other excited SU(6) multiplets\footnote{Note that
the suppression of two-body operators in the present case implies that
F-wave decay amplitudes should be highly suppressed.}.

Our second assumption is that mixing between states in the 
excited {\bf 56}-plet and other SU(6) multiplets, {\em i.e.} configuration 
mixing, can be neglected.  In the case of the {\bf 70}-plet,  successful 
large-$N_c$ effective theory descriptions of the strong decays~\cite{largen2}, 
photoproduction amplitudes~\cite{largen3}, and the nonstrange mass 
spectrum~\cite{largen1} all neglect mixing with other SU(6) multiplets.  
In the present case, one might worry that the nearest observed states 
with appropriate quantum numbers for mixing, members of the 
positive-parity {\bf 70}-plet, are only between 200 and 300 MeV heavier 
than the states of interest, so that a significant effect cannot be precluded. 
Fortunately, evidence from explicit quark models suggests that configuration 
mixing involving the excited {\bf 56} is also small~\cite{isgurkarl},
and even smaller than one would expect from $1/N_c$ arguments~\cite{buch}. 
We simply adopt this as a working assumption,  without wedding 
ourselves to a particular quark model or choice of baryon 
spatial wave function.

\section{Analysis of Observed States}

\begin{figure}
\vglue -1in  
\centerline{ \epsfxsize=6.5 in \epsfbox{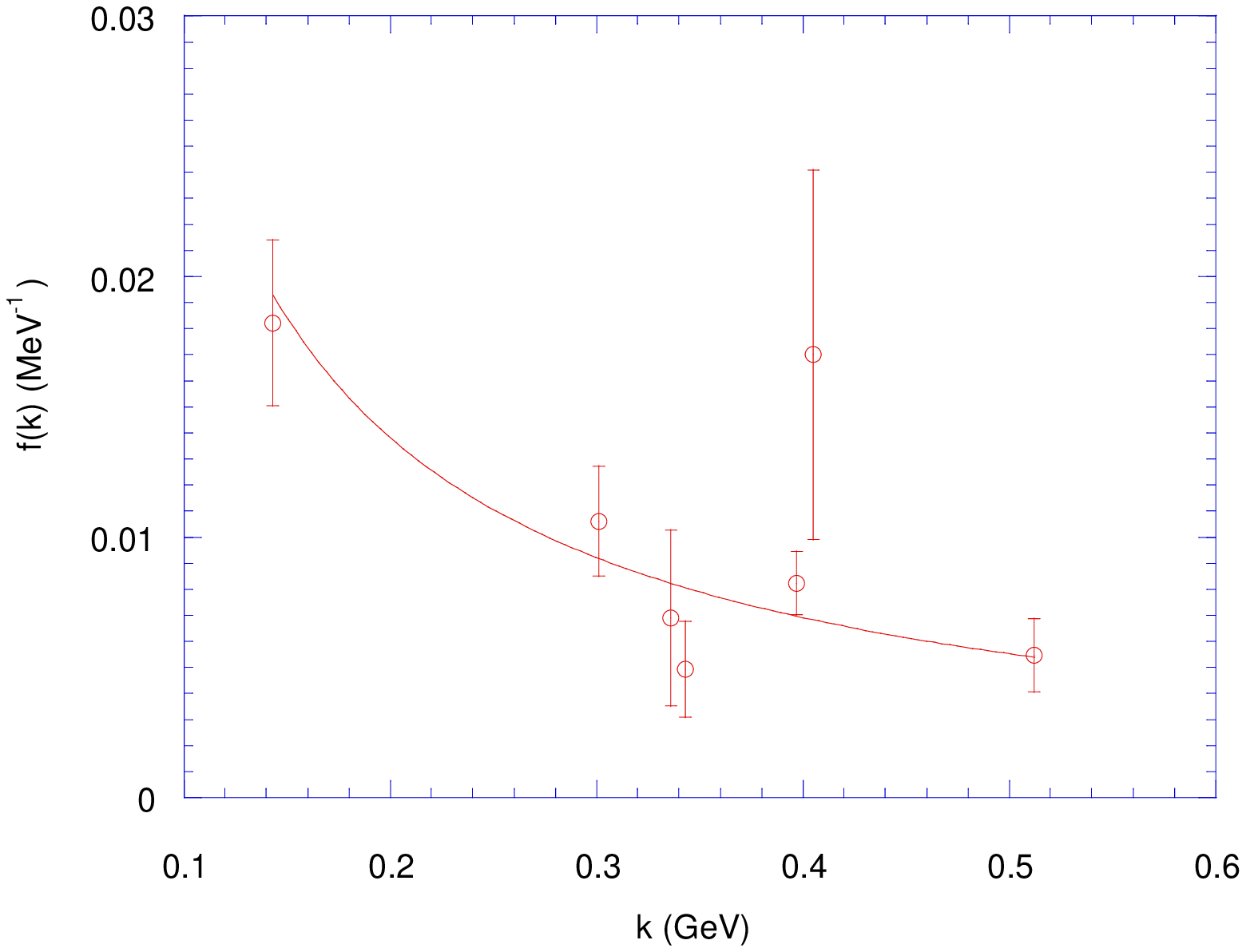}  }
\vglue -1in 
\caption{The function $f(k)$. The best fit corresponds
to $f=2.8/k$, with $\chi^2/{\rm d.o.f}=1.1$.}
\label{figure1}
\end{figure}

We first determine the functional form and normalization of $f(k)$
by considering the decay modes of observed members of the excited 
{\bf 56}-plet, the $N(1440)$, $\Delta(1600)$, $\Lambda(1600)$ and
the $\Sigma(1660)$.  We will refer to this multiplet as the 
{\bf 56$^\prime$} henceforth. (The RPP refers to the {\bf 56$^\prime$} 
as the $(56,0_2^+)$.) Our input values are shown in 
Table~\ref{table1}, and have been extracted from data in the 
Review of Particle Physics (RPP)~\cite{rpp}. The one 
standard deviation errors on the baryon masses were taken 
to be half of the corresponding mass ranges given in the RPP,
and partial widths were computed using the quoted full widths,
branching fractions, and associated errors.  The values for $f(k)$ 
corresponding to each observed decay mode are shown in Fig.~\ref{figure1}.  
Since we wish to model the function  $f(k)$ for momenta from $\sim 100$ 
to $\sim 500$~MeV, we require only that we find a simple functional
form that works well within this range.  The most successful result
is $f(k)=(2.8\pm 0.2)/k$, which has a $\chi^2$ per degree of
freedom of $1.1$.  Note that the functional forms $a+kb$, $a/k^{0.5}$,
$a/k^{1.5}$ lead to a $\chi^2$ per degree of freedom of $1.9$, $1.6$,
and $2.8$, respectively.  It is somewhat remarkable that such
a simple functional form can account for the data in the momentum
range of interest, though admittedly, the experimental uncertainties are 
large.  It is not inconceivable, for example, that improvement in the data 
could lead to a preference for the less singular-looking linear fit, but given
the present errors, the difference between a linear and the $1/k$ fit 
makes little difference in our decay predictions.  Continuing
in our spirit of model independence, we do not try to ascertain
the origin of $f(k)$, but explore how far we can go in predicting decay 
modes for unobserved states. 

\section{Mass Predictions}

We use the G\"ursey-Radicati (GR) formula to predict masses for the
unobserved members of the {\bf 56}-plet:
\begin{equation}
M=A+BN_s+C[I(I+1)-N_s^2/4]+DJ(J+1).
\label{eq:grf}
\end{equation}
Here J (I) is the baryon spin (isospin) and $N_s$ is the number of
valence strange quarks.  The GR formula predicts spin- and flavor-dependent 
mass splitting for completely symmetric SU(6) baryon multiplets in terms 
of the four parameters $A$, $B$, $C$ and $D$; in our case, these parameters
are determined from the masses given in Table~\ref{table1}.  While the 
GR formula was originally a conjecture~\cite{GR}, and later `derived' 
by assuming SU(6) symmetry and a specific set of mass operators transforming
in symmetry-breaking representations of small dimensionality~\cite{gr2}, we 
begin by showing how Eq.~(\ref{eq:grf}) may be obtained more rigorously 
from large-$N_c$ QCD.  This observation follows from the discussion in 
Ref.~\cite{manrev}, but is not presented there explicitly.  In a large-$N_c$ 
operator analysis for the {\bf 56}-plet, we may write the general mass formula
\begin{eqnarray}
M&=&a_1\openone + a_2 \frac{S^2}{N_c} + \epsilon a_3 T^8 + 
\epsilon a_4 \frac{S^i G^{i8}}{N_c} + +\epsilon a_5 \frac{S^2 T^8}{N_c^2}
+ \epsilon^2 a_6 \frac{T^8 T^8}{N_c} 
\nonumber \\
&+&\epsilon^2 a_7 \frac{T^8 S^i G^{i8}}{N_c^2}+\epsilon^3 a_8
\frac{T^8 T^8 T^8}{N_c^2}\, , 
\end{eqnarray}
where $S$, $T$, and $G$ are the spin, flavor, and spin-flavor generators
of SU(6) (with a sum over quarks left implicit), and 
$\epsilon \sim 1/3\sim 1/N_c$ parameterizes the size of SU(3) breaking.
Notice that the eight order-one coefficients, $a_1 \ldots a_8$, completely 
span the space of observables, namely the eight baryon mass eigenvalues.  
Assuming that the baryon states have spin, isospin, and strangeness of order 
one in the large $N_c$ limit, we may discard all but the first four terms 
if we choose to work only up to subleading order.  Acting on a large-$N_c$
baryon state
\begin{equation}
S^2=S(S+1) \, ,
\end{equation}
\begin{equation}
T^8 = (N_c-3N_s)/\sqrt{12} \, ,
\end{equation}
and for totally-symmetric spin-flavor wave functions
\begin{equation}
S^i G^{i8}=\frac{1}{4\sqrt{3}}\left[3I(I+1)-S(S+1)-3N_s(N_s+2)/4\right].
\end{equation}
These identities imply that the only effects of the discarded terms
up through order $1/N_c$ are redefinitions of the first
four coefficients, $a_1\ldots a_4 \rightarrow a'_1\ldots a'_4$.  
Thus we may write the mass eigenvalues
\begin{eqnarray}
M = a'_1 N_c &+& \frac{a'_2}{N_c} S(S+1) + \epsilon a'_3 (N_c-3N_s)
\nonumber \\ [1.5ex]
&+& \epsilon \frac{a'_4}{2\sqrt{12}N_c}
\left[3I(I+1)-S(S+1)-3N_s(N_s+2)/4\right] +{\cal O}(1/N_c^2) \,\,\, .
\end{eqnarray}
Eq.~(\ref{eq:grf}) then follows from a trivial redefinition of the 
coefficients. In modern language, the SU(6) breaking representations 
that lead to the GR mass formula are precisely those that arise at leading and
subleading order in the $1/N_c$ expansion.

Using the observed masses in Table~\ref{table1}, we may solve
for the coefficients in Eq.~(\ref{eq:grf}).  In units of MeV, we find
\begin{center}\begin{tabular}{ll}
$A=1406.3 \pm 31.3$, \qquad &
$B=195.0 \pm 43.0$, \\
$C=15.0 \pm 38.1$, \qquad &    
$D=43.3 \pm 46.0$ , 
\end{tabular}\end{center}
from which we predict the following states:
\begin{center}\begin{tabular}{llll}
$\Xi'(1825\pm 98),$ \qquad & ${\Sigma^*}'(1790\pm 192),$ \qquad &
${\Xi^*}'(1955\pm 196),$ \qquad & $\Omega'(2120\pm 234)$ .
\end{tabular}\end{center}

\section{Decay Width Predictions}

From Eq.~(\ref{eq:mop}), the decay width is given by
\begin{equation}
\Gamma = {M_f \over 6 \pi M_i} \, k^2 f(k)^2 \, | {\cal G} |^2 
\end{equation}
where $M_i$ and $M_f$ are the masses of the initial and final baryons and
\begin{equation}                    \label{MEsquared}
|{\cal G}|^2 \equiv 
     {\sum}  \  | \langle B_f | G_{ja} | B_i \rangle |^2 ,
\end{equation}
where the sum is over final state spins and isospins. The quantity
$|{\cal G}|^2$  may be obtained either by use of symbolic math 
code~\cite{largen2} or by using
\begin{equation}
\langle B_f | G_{ja} | B_i \rangle 
  = 
     \langle B_f (I_f,\alpha_f,S_f,m_f)
                 | G_{ja} | 
             B_i (I_i,\alpha_i,S_i,m_i) \rangle
  = {\cal G}
         \left(
                        \begin{array}{cc|c}
                          I_f      & I_a      & I_i \\
                          \alpha_f & \alpha_a & \alpha_i
                        \end{array}
                                          \right)
         \left(
                        \begin{array}{cc|c}
                          S_f      & 1      & S_i \\
                          m_f      & j      & m_i
                        \end{array}
                                          \right)
\end{equation}
and evaluating the left-hand side for one particular final state.
Here, $\alpha$ and $m$ stand for isospin and spin projections, and the sum
in equation~(\ref{MEsquared}) is over $\alpha_f$, $\alpha_a$, $m_f$, and
$j$.  

We have already discussed fitting the function $f(k)$ to the
measured decays, and how to use the G\"ursey-Radicati mass formula
to obtain the masses of the unobserved {\bf 56$'$} states.  Using the results
of the fit with $f(k) \propto 1/k$, we predict the remainder of the strong
decays of the {\bf 56$'$}.  The results are shown in Table~\ref{widths},
along with the masses we use for the initial baryons.

Table~\ref{widths} does not quote the uncertainties for the widths.  Numerical
uncertainties come from two sources.  One is the uncertainty in the
function $f(k)$, and the other is the uncertainty in the masses.  The
uncertainty in the fitted $f(k)$ leads to about $\pm 15$\% uncertainties
in the widths.  Further width uncertainty induced by mass uncertainty can
be large, as discussed below. Of course, once the state is found, the widths in
Table~\ref{widths} can be recalculated easily and accurately with the
correct mass. 

\section{Discussion}

The uncertainty in predicting the  width of the 
{\bf 56$^\prime$} decays is dominated not by uncertainty
in the matrix elements but by uncertainty in the masses of
the states.  Four of the {\bf 56$^\prime$} masses are
measured and four are not.  We must predict the
unmeasured masses, and do so using the G\"ursey-Radicati
mass formula. The accuracy of masses predicted
from the G\"ursey-Radicati formula is ultimately limited by
the the approximations inherent in its derivation, and is
easily estimated in a large-$N_c$ scheme.  Currently,
however, uncertainties in the predicted masses are dominated
by uncertainties in the measured masses, and they are not
small.

Table~\ref{widths} shows the predicted partial width of the various
decay modes, using the central values of the predicted masses. To help 
see how the mass uncertainties affect us, consider the equal spacing
rule that follows from the GR formula.  The decuplet spacing is
given by
\begin{equation}
\Delta m_{10'} = {3\over 2} m(\Lambda') -
     {1\over 2} m(\Sigma') - m(N')
   =  (165 \pm 110)  \ {\rm MeV}   .
\end{equation}
(For comparison, the equivalent prediction for the ground
state decuplet is $\Delta m_{10} = 139$ MeV, to be
compared to 153, 148, and 139 MeV for the 
$\Delta$-$\Sigma^*$,$\Sigma^*$-$\Xi^*$, and 
$\Xi^*$-$\Omega$ mass splittings, respectively.)  Then, as
one example, the predicted central value for the $\Omega'$
mass is 2.12 GeV, and this leads to a sizeable decay width
for $\Omega' \rightarrow \Xi^* \bar K$.  However, the
uncertainties allow the $\Omega'$ to lie below the threshold
for this decay.

Moreover, a simple weighted averaging of the published results that
the RPP~\cite{rpp} uses as the basis of their mass estimates for
the $\Lambda'$ and $\Sigma'$ gives somewhat different central
masses and much tighter uncertainties than they
conservatively quote. One finds

\begin{eqnarray}
m(\Lambda') = (1600 \pm 7) \ {\rm MeV} \nonumber  \\
m(\Sigma') = (1672 \pm 5) \ {\rm MeV}  .
\end{eqnarray}
(The error limit on the $\Lambda'$ mass includes a scale
factor $S = 1.4$, in accordance with procedures described in
the RPP narrative.)  Using these masses together with the
previous $N'$ and $\Delta'$ masses yields

\begin{equation}
\Delta m_{10'} = (114 \pm 20) \ {\rm MeV}
\end{equation}
This reduced spacing leads to a number of decays listed in 
Table~\ref{widths} being kinematically forbidden.

Unfortunately, there is not much one can do theoretically to reduce
the uncertainties originating from the {\bf 56$^\prime$} masses.  
Table~\ref{widths} simply presents the most reliable predictions we can 
make with the current data. As unobserved states are discovered, and their 
masses measured, one can easily revise the widths in Table~\ref{widths}. 
To reiterate, the relevant formula for our $kf(k) = const. = a$ fit is
\begin{equation}
\Gamma = {k M_f \over 6 \pi M_i}  \,  a^2  \, | {\cal G} |^2,
\end{equation}
where $M_i$, $M_f$, and $k$ are defined already, the matrix
element sum $|{\cal G}|^2$ is given in Table~\ref{widths}, and 
$a$ is 2.8 $\pm$ 0.20, based on the current data.

The function $f(k)$ is not taken from any model, but fit to the limited
data with the fulfilled hope that there would be a good fit using a
simple form.  The form one would expect from an atomic physics
calculation is rather different from the one we found.  For the
hydrogen atom, in a nonrelativistic calculation, $f(k)$ would be 
proportional to a matrix element of $\exp(i \vec k \cdot \vec r)$ 
between 2S and 1S wave functions.  Further in a hydrogen atom, the 
wavelength of the outgoing radiation is long compared to the Bohr radius, 
and the small $k$ result that $f(k) \propto k^2$ is valid.  This leads to 
a very small width and the famous metastability of the hydrogen 2S state.  
For particle decay, the wavelength of the outgoing meson is comparable to 
and often smaller than the size scale of the baryon states.  One should 
have no expectation that a low $k$ approximation will work.  This also 
applies to barrier factors in the non-metastable case. 

A model or an eventual {\it ab initio} calculation for the baryon states
will produce some definite $f(k)$ that may or may not look analytically
like our form, but to fit the data must not be numerically dissimilar to
our result. 

To summarize, we have without reference to models, but using the limited
available data, studied the strong decays of the lowest-lying radially 
excited baryons. We assumed, justified by previous studies, that one-body 
operators dominated and that configuration mixing could be neglected.  Simple 
forms for the one momentum-dependent function give a good fit to existing data,
and allow prediction of 22 additional decay modes for this multiplet.

\begin{center}
{\bf Acknowledgments}
\end{center}
We thank R. Horgan for useful communications, and R.~F.~Lebed for
a careful reading of the manuscript.  CDC and CEC thank the 
National Science Foundation for support under Grant No.\ PHY-9900657.  In 
addition, CDC thanks the National Science Foundation for support under 
grant No. PHY-9800741 and the Jeffress Memorial Trust for support under 
Grant No. J-532.


%
\begin{table}
\begin{center}
\begin{tabular}{ccccc}
          & Mass (MeV)    & Decay   & $k$ (MeV)  
& Partial Width (MeV)  \\ \hline
$N(1440)$ & $1450\pm 20$  &     &         & \\
          &               & $N\pi$ & $397$  & $227.5\pm 67.3$  \\
          &               & $\Delta \pi$ & $143$ & $87.5\pm 30.5$ \\ \hline
$\Delta(1600)$ & $1625\pm 75$ & & & \\
& & $N\pi$  & $512$ &$61.25\pm 31.5$ \\
& & $\Delta \pi$ & $301$ & $192.5\pm 76.0$ \\ \hline
$\Lambda(1600)$ & $1630\pm 70$ & & & \\
& & $N\overline{K}$ & $343$ & $33.75\pm 25.2$ \\
& & $\Sigma \pi$ & $336$ & $52.5\pm 51.3 $ \\ \hline
$\Sigma(1660)$ & $1660\pm 30$ & & &\\
& &  $N\overline{K}$ & $405$ &$24.0\pm 20.0$ 
\end{tabular}
\end{center}

\caption{Input values used in the determination of $f(k)$.}
\label{table1}
\end{table}
\begin{table}
\begin{center}
\tighten
\begin{tabular}{ccccc}
Decay                 & $M_i$(MeV) & $k$ (MeV)& $|{\cal G}|^2$ & 
 $\Gamma$ (MeV)  \\  \hline
$N' \rightarrow N \pi$ $^\dagger$ & 1450  &  397  & ${25/16}$  & 164
\\
$N' \rightarrow \Delta \pi$ $^\dagger$& 1450 &  143  & 2  & 99  
\\
$\Delta' \rightarrow N \pi$  $^\dagger$& 1625  & 512 & $1/2$ & 60  
\\
$\Delta' \rightarrow \Delta \pi$ $^\dagger$& 1625 & 301 & $25/16$ & 145 
\\
\hline
$\Lambda' \rightarrow N \bar K$ $^\dagger$& 1630 & 343 & $9/8$ & 91
\\
$\Lambda' \rightarrow \Sigma \pi$ $^\dagger$& 1630 & 336 & $3/4$ & 75 
\\
$\Lambda' \rightarrow \Sigma^* \pi$ & 1630 & 188 & $3/2$& 97
\\
\hline
$\Sigma' \rightarrow N \bar K$ $^\dagger$& 1660& 405& $1/24$& 4
\\
$\Sigma' \rightarrow \Lambda \pi$& 1660& 439& $1/4$& 30
\\
$\Sigma' \rightarrow \Sigma \pi$& 1660& 385& $2/3$& 75
\\
$\Sigma' \rightarrow \Sigma^* \pi$& 1660& 218& $1/3$& 25
\\
\hline
$\Xi' \rightarrow \Lambda \bar K$& 1825& 403& $1/16$& 6
\\
$\Xi' \rightarrow \Sigma \bar K$& 1825& 323& $25/16$& 134
\\
$\Xi' \rightarrow \Xi \pi$& 1825& 420& $1/16$& 8
\\
$\Xi' \rightarrow \Xi^* \pi$& 1825& 240& $1/2$& 41
\\
\hline
${\Sigma^*}' \rightarrow N \bar K$& 1790& 519& $1/6$& 18
\\
${\Sigma^*}' \rightarrow \Delta \bar K$& 1790& 214& $5/6$& 50
\\
${\Sigma^*}' \rightarrow \Lambda \pi$& 1790& 535& $1/4$& 34
\\
${\Sigma^*}' \rightarrow \Sigma \pi$& 1790& 484& $1/6$& 22
\\
${\Sigma^*}' \rightarrow \Sigma^* \pi$& 1790& 338& $5/6$& 89
\\
${\Sigma^*}' \rightarrow \Sigma \eta$& 1790& 196& $1/4$& 13
\\
\hline
${\Xi^*}' \rightarrow \Lambda \bar K$& 1955& 525& $1/4$& 31
\\
${\Xi^*}' \rightarrow \Sigma \bar K$& 1955& 461& $1/4$& 29
\\
${\Xi^*}' \rightarrow \Sigma^* \bar K$& 1955& 239& $5/4$& 86
\\
${\Xi^*}' \rightarrow \Xi \pi$& 1955& 520& $1/4$& 36
\\
${\Xi^*}' \rightarrow \Xi^* \pi$& 1955& 358& $5/16$& 36
\\
${\Xi^*}' \rightarrow \Xi \eta$& 1955& 269& $1/4$& 19
\\
\hline
$\Omega' \rightarrow \Xi \bar K$& 2120& 506& 1& 128
\\
$\Omega' \rightarrow \Xi^* \bar K$& 2120& 274& 5/4& 101
\\
\end{tabular}

\end{center}

\caption{Decays of the radially excited {\bf 56$'$} into ground state
{\bf 56} plus meson.  Seven of these decays are measured (indicated
by a dagger), and the other 22 widths are predictions.  If the mass of
the initial state is known, the uncertainty in the width
from our fitting procedure is $\pm 15$\%.  Uncertainties in the
mass induce further and sometimes large uncertainties in the width.
These of course are greatly reducible once the mass of the state is
measured, as discussed in the text.}

              \label{widths} 

\end{table}
\end{document}